# Energy transfer reaction K(4s) + K(7s) → K(4s) + K(5f), theory compared with experiment

M. Głódź[a*], A. Huzandrov[a], S. Magnier[b], L. Petrov[ca], I. Sydoryk[a], J. Szonert[a], J. Klavins[d], K. Kowalski[a]

[a] *Institute of Physics, Polish Academy of Sciences, Al. Lotników 32/46, 02-668 Warsaw, Poland*

[b] *Univ. Lille, CNRS, UMR 8523- PhLAM- Laboratoire de Physique des Lasers, Atomes et Molécules, F-59000 Lille, France*

[c] *Institute of Electronics, Bulgarian Academy of Sciences, 1784 Sofia; Boul. Tsarigradsko Shosse 72, Bulgaria*

[d] *Institute of Atomic Physics and Spectroscopy, University of Latvia, 1586 Riga, Latvia*

**Abstract**

A comparison between theory and experiment, concerning the K(4$s$)+K(7$s$)→K(4$s$)+K(5$f$) reaction of energy transfer in thermal collisions, is presented. Relevant cross sections are calculated for the potassium vapour temperatures in the range of 310-1000 K. They are based on the theoretical adiabatic $K_2$ potential energy curves and on the use of the multicrossing Landau-Zener model. In the temperature range of the present experiment, 428-451 K, the computed cross sections vary little, from $2.09 \times 10^{-14}$ cm$^2$ to $2.04 \times 10^{-14}$ cm$^2$, and agree well with the value $1.8(8) \times 10^{-14}$ cm$^2$, which is the average of the corresponding experimental results.

---

[*] Corresponding author: Fax: +48 22 843 09 26  *E-mail address*:glodz@ifpan.edu.pl (M. Głódź)

## 1. Introduction

We studied the process of excitation energy transfer (ET) in thermal vapour-cell collisions of the excited potassium atom K(7*s*) with the parent ground state atom K(4*s*), for the case of the reaction

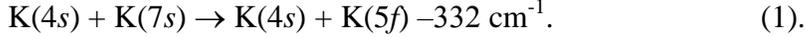
$$K(4s) + K(7s) \to K(4s) + K(5f) - 332 \text{ cm}^{-1}. \quad (1).$$

Reaction (1) is an implementation of the following reaction

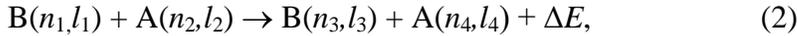
$$B(n_1, l_1) + A(n_2, l_2) \to B(n_3, l_3) + A(n_4, l_4) + \Delta E, \quad (2)$$

which, in its general form, symbolizes the process of ET in binary collisions of an atom B with an atom A. A and B may be atoms of the same or different species. $\Delta E$ is the thermal energy compensating for either excess- or defect- energy-difference

$$\Delta E = \left(E_{B(n_3,l_3)} + E_{A(n_4,l_4)}\right) - \left(E_{B(n_1,l_1)} + E_{A(n_2,l_2)}\right).$$

We start by listing some of the fields in which atomic collisions are of interest. We have particularly in mind inelastic binary collisions and alkali metal atoms as collisional partners.

Most obvious fields are: spectroscopy of atoms in a gas phase including determination of atomic states natural-lifetimes, the study of kinetics related to chemical processes, the laboratory-plasma physics. This also applies to astrophysics of cosmic plasmas [1-3]. It is worth to note that among other elements, also alkali metal atoms (including potassium atoms) play a role in cosmic collisional encounters.

Characteristics of atomic collisions are also crucial in some branches of technology in order to asses either a positive or negative impact of collision initiated processes. For example, in the field of laser construction projects, the operation principles of the *collision lasers operating on atomic transitions*, are directly founded on ET collisions [4], while the development of the high-power *diode-pumped alkali vapour lasers* [5] may suffer from an ET process, which is energy pooling (EP) [6,7].

Diverse and quite advanced is the theme of cold collisions in cooled and trapped samples of atoms, often alkali atoms, (also of cold atoms excited to Rydberg states [8,9]) at temperatures $T < 1$ mK, often classified as "ultracold" temperatures, as *e.g.*, in Ref.[10]). Ultracold collisions in light field result in trap loss but also, for instance, in formation of ultracold molecules (*via* photoassociation) for high resolution molecular spectroscopy, or in photoassociative ionization [10-13].

In the regime of ultracold temperatures, the role of collisions is investigated down to mikrokelvins and nanokelvins, in which temperatures quantum degeneracy occurs [12].

Despite all peculiarities of ultracold collisions, in both regimes of "hot" and ultracold collisions, the collisional process can be often described in terms of a quasi-molecule- or molecule-formation, and it is associated with the properties of the potential energy curves (PECs) of the electronic states. Therefore, we believe that the development in theoretical descriptions of this kind, tested experimentally with respect to collisions at thermal energies (like in our present work), may valuably serve both areas. Alkali metal atoms, as characterized with single valence electron, are most often used in such investigations.

In the following, we concentrate entirely on inelastic hot thermal alkali-alkali collisions, and not, for instance, on such alkali-noble gas collisions. The latter, leading to population quenching, are discussed in Chapter 5.5.1 in [14].

In the last few decades, a lot of work has been done on reactions of energy pooling (EP)[1], mentioned above, (and perhaps less on reverse energy pooling), also with participation of potassium, *e.g.*, [7,15-18].

Less often there are studied reactions of ET in collisions of a ground state atom (*e.g.*, A in reaction (2)) with an excited atom (B in (2)). This particularly refers to atoms B in an f-state (*n*, *l*=3) on either right- or left-hand side of reaction (2) with exception of widely investigated specific kind of ET, which is the quasi-elastic *l*-mixing between the nearly degenerate alkali states of the same hydrogen-like manifold, *e.g.*, see Chapter 5.5.2 in [14] and [19] (the latter reference concerns an experiment with K atoms).

To the best of our knowledge, reaction (1), the subject of this article, has not been studied before neither theoretically, nor experimentally. In fact, for potassium K-K* collisions it is even hard to find in the literature reports on other analogous reactions. An experiment concerning the reaction

---

[1] The EP reaction starts with two colliding atoms (A and B in reaction (2)), both initially in excited states, and ends up with one atom in a highly excited state, and the other in the ground state [7].

$$K(4s) + K(6s) \rightarrow K(4s) + K(4d) + 53 \text{ cm}^{-1} \qquad (3)$$

was reported in Ref. [20] and the determined cross section (relatively small) was discussed in the review of $s \rightarrow d$ transfer reactions [21].

One of the aims of the present work is to contribute at filling gaps of lacking characteristics of K-K* collisions, and specifically to target the case of a considerable difference in angular momentum quantum numbers $\Delta l=3$ between the final $f$-state and the initial $s$-state.

From the theoretical perspective, it is rare to reach so high alkali excited states (as are $f$ states) in calculations, because often only the two first asymptotes, $s+s$ and $s+p$, are needed in calculations concerning cold molecules or transfer reactions. Usually, people look at molecular states located below the $p+p$ limit and, in particular, at those dissociating into $ns+np$. In view of this, calculated cross sections for a collisional process involving $f$ states, confronted with measured ones, are a useful test to check the accuracy of the relevant molecular theoretical data, since spectroscopy of molecular states correlated with $f$ states is extremely rare.

In this article we focus on a theoretical interpretation of the process of collisional reaction (1). The theoretical approach and the results of calculations are presented in Section 2. Our experiment is briefly described in Section 3. More details concerning the experimental procedures, data acquisition and processing can be found in Reference [22]. The theoretical results are compared with the experimental ones in Section 4.

## 2. Theoretical

Potential energy curves (PECs) involved in the interpretation of the excitation energy transfer reaction (1) have been determined in the framework of a pseudopotential method, in a similar way as in Ref. [23]. One $s$ and one $p$ Gaussian-type orbitals have been added to the basis set used in Ref. [23] in view to describe the $K(4s)+K(7s)$ and $K(4s)+K(5f)$ asymptotes. As previously, a satisfying agreement with experimental data listed in Ref. [23] has been obtained both for adiabatic potential energy curves and spectroscopic constants. The averaged differences on spectroscopic constants for 25 molecular states are found to be $\Delta R_e = 0.06 a_0$, $\Delta \omega_e = 0.75 \text{cm}^{-1}$, $\Delta T_e = 68 \text{cm}^{-1}$ and $\Delta D_e = 79 \text{cm}^{-1}$.

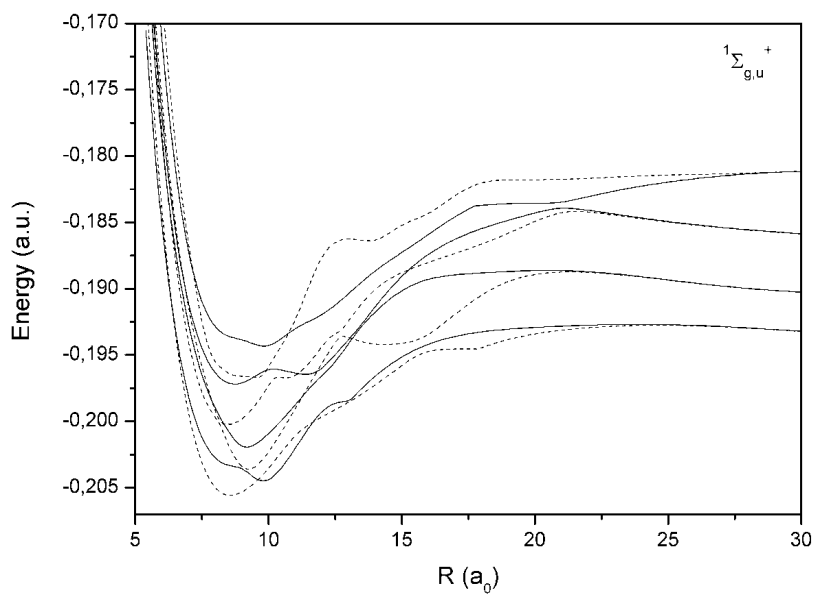

**Figure 1.** Potential energy curves of $^1\Sigma_g^+$ (solid line) and $^1\Sigma_u^+$ (dashed line) electronic states dissociating into K(4*s*)+K(6*p*) and up to K(4*s*)+K(5*f*). Energy values and internuclear distances are given in atomic units.

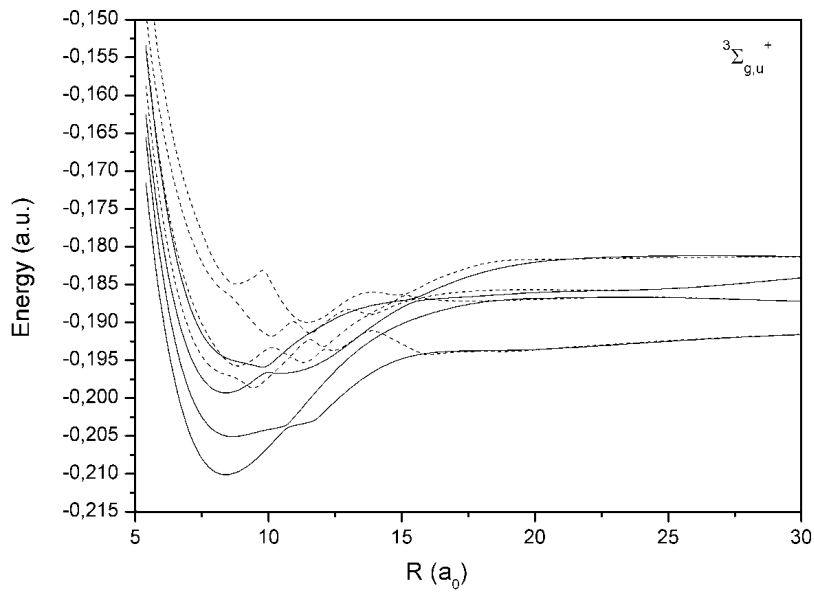

**Figure 2.** Potential energy curves of $^3\Sigma_g^+$ (solid line) and $^3\Sigma_u^+$ (dashed line) electronic states dissociating into K(4*s*)+K(6*p*) and up to K(4*s*)+K(5*f*). Energy values and internuclear distances are given in atomic units.

Theoretical cross sections for energy transfer reaction (1) have been calculated by applying a semi classical multicrossing Landau-Zener model [24] used successfully in the treatment of energy pooling reactions between sodium and potassium atoms [16,17]. Calculations are based on the estimation of the population transfer from one PEC involved in the process to another one in the vicinity of each avoided crossing labelled by $n$ in the text below. PECs for $^{1,3}\Sigma_{g,u}^{+}$ molecular states correlated to the asymptotes K(4$s$)+K(6$p$) and up to K(4$s$)+K(5$f$) have been considered. They are displayed in Figures 1 and 2. Numerous avoided crossings are present, in particular, at intermediate internuclear distance. They are partly due to the ionic-covalent interaction, the ionic state being in the ground state or in an excited state. Landau-Zener parameters (position $R_n$, diabatic potential energy $U_n$, splitting between the two PECs $V_n$, difference of the slopes $\Delta F_n$ at $R_n$) have been extracted from the present potential energy curves for each avoided crossing. They are listed in Table I, and the value $P_n$ of Landau-Zener probability of transition between two curves at the avoided crossing $n$ is given for one value of impact parameter ($b$=0) and one collisional energy ($E = 400\,\mathrm{K}$). Except for avoided crossings located at $14.48 a_0$ and $21.44 a_0$ in the $^1\Sigma_u^+$ PECs correlated to $\mathrm{K}(4s) + \mathrm{K}(7s)$ and $\mathrm{K}(4s) + \mathrm{K}(5f)$, all avoided crossings contribute to the process. The probabilities are estimated to be $\approx 0.900$ for several impact parameters and collisional energies, which leads to a nearly full population transfer from one curve to the other at the corresponding avoided crossing.

Total cross sections $\sigma(\upsilon)$ for each molecular symmetry have been determined by integrating numerically the final population of each possible exit channel correlated with 4$s$+5$f$ over different impact parameters. These values are still dependent on the centre-of-mass collision energy and therefore on the relative velocity $\upsilon$ of the colliding partners. Having in mind the vapour-cell experiment, in which thermally averaged collisional transfer rates are measured (see Sect. 3), the relevant thermally averaged theoretical cross sections have been calculated. They are defined as

$$\overline{\sigma}(T) = \langle \sigma(\upsilon)\upsilon \rangle / \overline{\upsilon}, \qquad (4)$$

where $\langle \sigma(\upsilon)\upsilon \rangle = k$ is the thermally averaged rate constant, corresponding to a certain vapour temperature $T$. The conventional Maxwell distribution over relative collision velocities has been

**Table I:** Landau-Zener parameters ($R_n, U_n, V_n, \Delta F_n$) of each avoided crossing $n$ present in the $^{1,3}\Sigma^+_{g,u}$ potential energy curves involved in the energy transfer reaction $K(4s) + K(7s) \to K(4s) + K(5f)$. Probability value ($P_n(b,E) = \exp(-2\pi \frac{|V_n|^2}{\upsilon_{R_n}(b,E)\Delta F_n})$ with $\upsilon_{R_n}(b,E) = \sqrt{\frac{2E}{\mu}(1 - \frac{b^2}{R_n^2} - \frac{U_n}{E})}$) is indicated for $b = 0$ and $E = 400K$. Parameter values are given in atomic units.

| $n$ | States | $R_n$ | $U_n$ | $V_n$ | $\Delta F_n$ | $P_n(0, 400K)$ |
|---|---|---|---|---|---|---|
| $^1\Sigma^+_g$ | | | | | | |
| 1 | $4s5d - 4s7s$ | 06.66 | -7.10x10$^{-3}$ | 7.14x10$^{-4}$ | 1.64x10$^{-2}$ | 0.942 |
| 2 | $4s5d - 4s7s$ | 13.44 | -1.16x10$^{-2}$ | 7.82x10$^{-5}$ | 2.54x10$^{-3}$ | 0.995 |
| 3 | $4s7s - 4s5f$ | 20.82 | -2.47x10$^{-3}$ | 2.17x10$^{-4}$ | 6.06x10$^{-4}$ | 0.859 |
| $^1\Sigma^+_u$ | | | | | | |
| 1 | $4s6p - 4s5d$ | 09.65 | -2.25x10$^{-2}$ | 5.74x10$^{-4}$ | 4.15x10$^{-3}$ | 0.863 |
| 2 | $4s6p - 4s5d$ | 15.69 | -1.31x10$^{-2}$ | 5.87x10$^{-4}$ | 1.15x10$^{-3}$ | 0.565 |
| 3 | $4s5d - 4s7s$ | 08.15 | -1.87x10$^{-2}$ | 1.23x10$^{-4}$ | 3.03x10$^{-3}$ | 0.991 |
| 4 | $4s5d - 4s7s$ | 12.66 | -1.22x10$^{-2}$ | 2.13x10$^{-4}$ | 7.27x10$^{-3}$ | 0.985 |
| 5 | $4s7s - 4s5f$ | 09.93 | -1.55x10$^{-2}$ | 7.00x10$^{-4}$ | 9.09x10$^{-3}$ | 0.903 |
| 6 | $4s7s - 4s5f$ | 14.48 | -6.34x10$^{-3}$ | 1.73x10$^{-3}$ | 3.62x10$^{-3}$ | 0.200 |
| 7 | $4s7s - 4s5f$ | 21.44 | -1.68x10$^{-3}$ | 1.23x10$^{-3}$ | 5.68x10$^{-4}$ | 0.005 |
| $^3\Sigma^+_g$ | | | | | | |
| 1 | $4s6p - 4s5d$ | 10.67 | -2.25x10$^{-2}$ | 2.23x10$^{-4}$ | 3.00x10$^{-3}$ | 0.969 |
| 2 | $4s5d - 4s7s$ | 13.52 | -1.11x10$^{-2}$ | 8.79x10$^{-4}$ | 2.57x10$^{-2}$ | 0.944 |
| 3 | $4s7s - 4s5f$ | 09.88 | -1.50x10$^{-2}$ | 3.99x10$^{-4}$ | 3.00x10$^{-3}$ | 0.903 |
| 4 | $4s7s - 4s5f$ | 15.48 | -5.78x10$^{-2}$ | 2.43x10$^{-4}$ | 1.58x10$^{-3}$ | 0.929 |
| $^3\Sigma^+_u$ | | | | | | |
| 1 | $4s6p - 4s5d$ | 10.81 | -1.32x10$^{-2}$ | 9.32x10$^{-5}$ | 2.81x10$^{-3}$ | 0.994 |
| 2 | $4s6p - 4s5d$ | 12.09 | -1.21x10$^{-2}$ | 1.90x10$^{-4}$ | 2.31x10$^{-3}$ | 0.971 |
| 3 | $4s5d - 4s7s$ | 11.57 | -1.06x10$^{-2}$ | 3.96x10$^{-4}$ | 2.84x10$^{-3}$ | 0.900 |
| 4 | $4s5d - 4s7s$ | 13.79 | -7.66x10$^{-3}$ | 9.18x10$^{-5}$ | 1.78x10$^{-3}$ | 0.991 |
| 5 | $4s5d - 4s7s$ | 15.74 | -5.49x10$^{-3}$ | 4.56x10$^{-5}$ | 7.43x10$^{-3}$ | 0.995 |
| 6 | $4s7s - 4s5f$ | 11.06 | -8.53x10$^{-3}$ | 1.00x10$^{-4}$ | 1.18x10$^{-3}$ | 0.984 |

used here. $\bar{\upsilon}$ is the mean relative velocity of K atoms at temperature $T$

$$\bar{\upsilon} = \sqrt{8k_B T/\pi\mu} \tag{5}$$

with $\mu$ - their reduced mass, and $k_B$ - the Boltzmann constant. From here on we abbreviate $\bar{\sigma}(T) \equiv \bar{\sigma}$.

Sum and contributions of each molecular symmetry are displayed in Fig. 3 for the thermally averaged cross sections. The major contribution comes from avoided crossings of $^3\Sigma_g^+$ and $^3\Sigma_u^+$ molecular states.

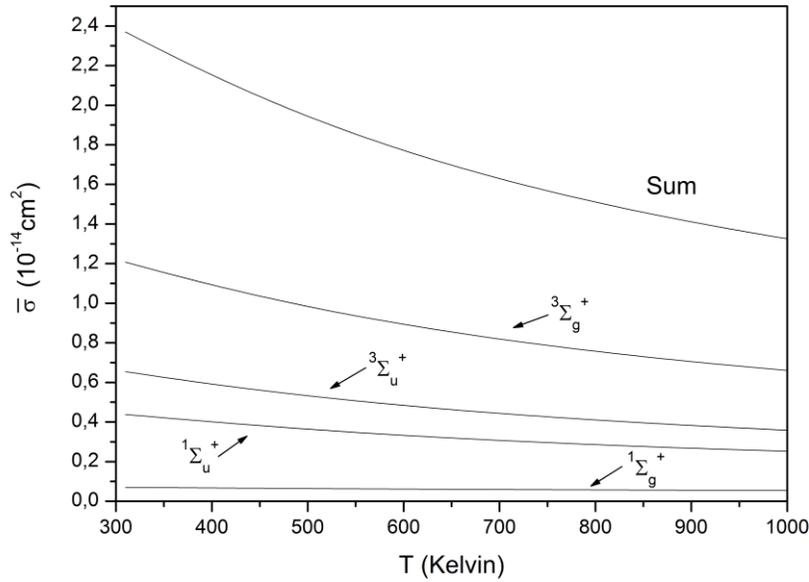

**Figure 3.** Variation of the thermally averaged cross sections for the $K(4s) + K(7s) \rightarrow K(4s) + K(5f)$ reaction as a function of the centre of mass collision energy expressed in Kelvins.

Comparison of the theoretical results with the experimental ones obtained for a range of temperatures centred at about 440 K is very satisfying, as presented in Sect. 4.

3. **Experimental**

The potassium atoms in a glass vapour-cell were directly excited to the 7s state by short laser pulses ($\lambda$=660.4 nm) in the two-photon 4s→7s transition. A laser with a few ns pulse duration and a few GHz linewidth was used. The development in time of the direct-fluorescence $I_{7s}$ and of the collisionally sensitized fluorescence $I_{5f}$ was registered sideways with respect to the direction of the laser beam, on the respective transitions: 7s→4p ($\lambda_{7s}$=579.2 nm) and 5f→3d ($\lambda_{5f}$=1102.1 nm). The upper limit for the laser-beam intensity in the detection zone in the cell, applicable for the experiment, was established. By rising intensity above a certain value (this limit), the values determined for cross sections became intensity-dependent, which indicated the break-down of the adopted procedure(s) of the experimental-data processing. This coincided with the observation that at intensities above the upper limit, some lines, not seen below it, appeared in the additionally registered fluorescence spectra (see below).

A monochromator with widely open slits (to minimise the escape of atoms out of the viewing range), a fast-response photomultiplier with S-1 photocathode and a multichannel scaler for photon-counting (with time resolution 5ns/channel) constituted the main components of the detection system. The measurements were carried out for different values of potassium vapour temperature varied in the range of 428-451 K, which corresponded to the range $9.2 \times 10^{12}$-$3.5 \times 10^{13}$ cm$^{-3}$ of the number density $N$ of potassium atoms.

By assuming that the only significant processes in our experiment are the direct 7s→5f collisional transfer (with the rate $R_{7s \to 5f}$), and the total decays of the 7s and 5f states' populations $N_i$ (with the respective rates $\Gamma_{7s}$ and $\Gamma_{5f}$), the two rate-equation model represents the evolution of these populations (see, for instance, equations (6) in Ref. [25]). In the framework of such a model, which we adopted as a starting point in data processing, the levels other than 7s and 5f were taken into account only in the role of a sink with respect to the total depopulation of each of the two states of interest. In this model (as well as in the finally used extended one), no back-stream of the population transfer was assumed; this seems well-grounded given the conditions of our experiment. Solutions of the two model equations, under the initial conditions: $N_{7s}(t=0)=N_{7s}^0$ and $N_{5f}(t=0)=0$, are a single-exponential and a double-exponential functions for time dependence of $N_{7s}$ and $N_{5f}$, respectively.

With these solutions and by applying the relation $I_i = A_{i-k} N_i$, where $A_{ik}$ denotes the spontaneous emission rate for the $i \to k$ transition, one obtains the following functions

$$I_{7s} = I_{7s}^0 \exp(-\Gamma_{7s} t), \tag{6a}$$

$$I_{5f} = \xi \frac{A_{5f-3d}}{A_{7s-4p}} I_{7s}^0 \frac{R_{7s \to 5f}}{\Gamma_{5f} - \Gamma_{7s}} \left( e^{-\Gamma_{7s} t} - e^{-\Gamma_{5f} t} \right) \tag{6b}$$

for time dependence of the direct-fluorescence intensity $I_{7s}$ and the collisionally sensitized-fluorescence intensity $I_{5f}$, respectively, to be fitted to the corresponding experimentally registered signals. The factor $\xi$ in (6b) was additionally introduced to compensate for different spectral sensitivities of the detection system at $\lambda_{5f}$ and $\lambda_{7s}$ wavelengths, and $\xi$ was determined *via* spectral calibration of the system. Other experimental conditions were preserved during the entire experiment, the same for both signals $I_{7s}$ and $I_{5f}$ of each pair registered in the measurements differing by temperature. Measures were also taken to compensate for (small) noise background in the experimental signals.

For a given measurement, the initial amplitude $I_{7s}^0 = I_{7s}(t = 0)$ and the value for the rate $\Gamma_{7s}$ were obtained by fitting the single-exponential function (6a) to experimental $I_{7s}$ signals. In fitting, we cut off few initial channels in which a short peak of a very fast decaying fluorescence was observed. Past the peak, the $I_{7s}$ signal stabilized to a single exponential dependence. Values for $\Gamma_{5f}$ were taken from our other experiment [26]. Please note, that to both of these measured ("effective") decay rates $\Gamma_i$, besides the spontaneous-emission rate (inverse of state's natural lifetime), two other components additively contributed. One was due to transitions induced by thermal radiation (BBR from "blackbody radiation") present in the cell, and the other due to all collisional-transfer processes from state *i* (collisional quenching). Both components are temperature-dependent, thus so are $\Gamma_i$. For $A_{5f-3d}$ and $A_{7s-4p}$, values recommended by NIST [27] were taken. Consequently, in this approach, the transfer rate $R_{7s \to 5f}$ was the only free parameter in fitting Formula (6b) to the experimental sensitized-fluorescence signal.

The experimental cross section $\sigma_{7s \to 5f}$ is defined as

$$\sigma_{7s \to 5f} = R_{7s \to 5f} / N\bar{v}, \tag{7}$$

and the rate constant $k_{7s \to 5f}$ is bound to $R_{7s \to 5f}$ as

$$k_{7s \to 5f} = R_{7s \to 5f} / N, \qquad (8)$$

where *N* is the number density of the perturbing potassium atoms in the ground state. Expecting the nearly Maxwellian distribution of relative collision velocities of potassium atoms in our vapour cell, for a given temperature *T*, there is a correspondence between the theoretical rate constant $k = \langle \sigma(\upsilon)\upsilon \rangle$ (see above, explanation to Formula (4)), and the experimental rate constant $k_{7s \to 5f}$, thus also between theoretical cross section $\overline{\sigma}(T) \equiv \overline{\sigma}$, defined by (4), and experimental cross section $\sigma_{7s \to 5f}$, defined by (7), where for $\overline{\upsilon}$ Formula (5) is taken.

When performing fitting procedure, it turned out that to get a more precise reconstruction of the time dependence of the registered $I_{5f}$ signals, it was necessary to compensate for some feeding of the 5f state population from sources in addition to the 7s→5f transfer considered above. In order to identify which levels and processes should be taken into account as meaningful, supplementary measurements were performed: (*i*) Emission spectra were registered of fluorescence photons being simultaneously cumulated within two time gates in two channels of the photon counter (one gate opened for the period from 0 to 50 ns, and the other from 50 ns to usually 10 μs), in a possibly broad spectral range (with our system we were able to register neither far IR, nor far UV lines). (*ii*) From some levels, lower than 7s, the time resolved fluorescence was registered with 5 ns resolution, in the same way as that from 7s or 5f.

By analyzing the experimental material, and performing various estimates (*e.g.,* concerning transitions with possible involvement of BBR, which appeared negligible), we found that two additional transfer channels had to be taken into account.

One of these channels was related to the observed, but neglected in the first approach, sharp peak in the 7s fluorescence about *t*=0. With limited time–resolution of our system we were unable to establish its actual time dependence. To make the fitting procedure simple, we decided to treat the peaked fluorescence as if it originated from an "additional state X" which contributed to ET, and we adopted a rough approximation that its population also decays exponentially, but much faster than population of the 7s state.

Most probably, this "peak" in sideways observed fluorescence from 7*s* is an evidence that the laser-excited population of 7*s* becomes fastly depleted as a result of a short pulse of, for instance, superradiance emitted at t≈0 along the laser beam. A superradiant cascading might develop on transitions down from 7*s*. Fast cascading of this kind has been for decades observed in alkalies, see *e.g.*, Refs [28,29]. We were unable to directly observe pulses of superadiance, but initial fragments of decaying fluorescence signals registered in measurements (*ii*) from some states lower than 7s, seem to confirm the existence of cascades in our experiment. In the case of fast cascading, especially if cascades would develop into multiple branches, several atomic states lower than 7s (including those, fluorescence from which we were unable to register) may become populated shortly after the exciting laser pulse, and pulses of photons at various wavelengths, released in cascades along the laser beam, are available for absorption in some coinciding transitions, *e.g.*, on the wings of some atomic lines. A population of excited molecular states could be expected as well. If it is followed by fast dissociation, the excited atoms can be also created in this way at about t=0. Analyzing the emission-spectra (in both time channels in measurements (*i*)), we were particularly careful in searching for lines from *d*-states, from which the 5*f* state could be populated in allowed radiative transitions. We have found that such channel might exist from the 6*d*-state (lines from other *d*-states were not present in the spectra), populated in the first 50 ns after *t*=0, *e.g., via* excitation of fast dissociating molecular states correlated with the asymptote 4s+6d. Therefore, we assumed a single exponential decay of 6*d* population with the effective $\Gamma_{6d}$ rate. To account for the process of collisional ET $X \rightarrow 5f$ and for population of 5*f* in $6d \rightarrow 5f$ spontaneous emission, two respective double-exponential expressions,

$I_{X \rightarrow 5f} = \alpha_X \left( e^{-\Gamma_X t} - e^{-\Gamma_{5f} t} \right) / \left( \Gamma_{5f} - \Gamma_X \right)$ and $I_{6d \rightarrow 5f} = \alpha_{6d} \left( e^{-\Gamma_{6d} t} - e^{-\Gamma_{5f} t} \right) / \left( \Gamma_{5f} - \Gamma_{6d} \right)$, were

added to (6b). Such three-component formula was fitted to registered $I_{5f}$ signals. In addition to $R_{7s \rightarrow 5f}$, also proportionality factors $\alpha_X$ and $\alpha_{6d}$, as well as $\Gamma_X$ were treated as free parameters of the fit. The fitted $\Gamma_X$ values approximately agreed with those expected from the duration of initial peaks on the registered $I_{7s}$ fluorescence signals.

The $\Gamma_{6d}$ parameter was either fixed on the values of effective rates $\Gamma_{6d}$, determined in our other (unpublished) experiment performed under similar conditions, or it was taken as a free parameter of the fit. In the latter case, values for the parameter $\Gamma_{6d}$ were obtained, on average, slightly different from those of the experiment, however the values for $R_{7s \to 5f}$ obtained with $\Gamma_{6d}$ fixed resp. free, differed very little (by few percent).

In Section 4, the refined values for $\sigma_{7s \to 5f}$, obtained in the second approach, are presented for the experimental results. However, the $\sigma_{7s \to 5f}$ values obtained in two approaches differ on the average by much less than the estimated experimental error. More details concerning the supplementary measurements (as well as the main one), data acquisition and data processing will be available in Ref. [22], under preparation.

### 4. Results and conclusions

In Fig. 4, the theoretical $\overline{\sigma}$ results and experimental $\sigma_{7s \to 5f}$ results are jointly displayed as $\sigma$ values dependent on temperature *T*. The main figure is the plot, denoted by "Sum" in Fig. 3. The small grey strip over *T*-axis marks the range of temperatures for which the experiment was performed. In the inset, the theoretical results in this range are compared with the corresponding experimental ones.

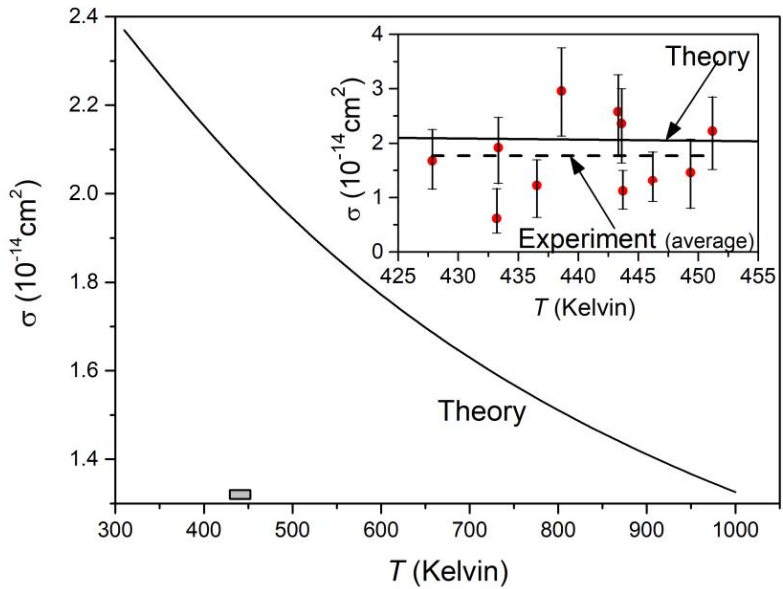

**Figure. 4.** *Main plot:* Calculated thermally averaged cross sections, sum of contributions of each molecular symmetry ("Sum" in Fig. 3) as a function of temperature *T*. The range of *T* of the experiment is marked in gray over the *T*-axis. *Inset:* The theoretical predictions in this range (full line) compared with the experimental values (points). The dashed line marks the average of the experimental values.

In the relatively small range of temperatures of the experiment the cross section values are expected to be nearly constant; the theory predicts only about 2.5% decrease with rising *T*, which is much less than the estimated uncertainty limits marked with each of the experimental values. Therefore for the representative experimental result in this range we have taken the average, amounting to $1.8(8) \times 10^{-14}$ cm$^2$. This value is in a very satisfying agreement with the theoretical ones spanned from $2.09 \times 10^{-14}$ cm$^2$ to $2.04 \times 10^{-14}$ cm$^2$.

In conclusion, we have calculated the thermally averaged cross sections for the excitation energy transfer reaction $K(4s) + K(7s) \rightarrow K(4s) + K(5f)$ for the range of potassium vapour temperatures of 310-1000 K. For the range of the temperatures of our experiment 428-451 K, the theoretical cross section values well reproduce the experimental ones. This agreement has allowed us to check, and then validate, the accuracy of potential energy curves of these high excited molecular states for which experimental spectroscopy remains unknown.

Since the experiment was performed with resolution in time, and supplemented with additional measurements, we were able to trace the competing processes that accompany the process of interest in this work, and to compensate for them in the course of data processing.